\documentclass[%
 aip,
 jmp,%
 amsmath,amssymb,
 reprint,%
]{revtex4-2}

\usepackage{graphicx}% Include figure files
\usepackage{dcolumn}% Align table columns on decimal point
\usepackage{bm}% bold math
\usepackage{chemformula} % Formula subscripts using \ch{}
\usepackage[T1]{fontenc} % Use modern font encodings
\usepackage{todonotes}
\usepackage{lineno}
\usepackage{amsmath,amsfonts,amssymb}
\usepackage[colorlinks=true,citecolor=blue,allcolors=blue]{hyperref}
\usepackage[caption=false,font=normalsize]{subfig}
\usepackage{lmodern}
\usepackage[T1]{fontenc}
\usepackage{contour}
\usepackage{textcomp}
\usepackage{epstopdf}

\definecolor{originred}{HTML}{F47575}
\definecolor{origindarkgrey}{HTML}{515151}
\definecolor{originblue}{HTML}{1A6FDF}
\definecolor{originlightblue}{HTML}{ECF4FD}
\definecolor{originlightred}{HTML}{FDEAEA}
\definecolor{originorange}{HTML}{FF8000}
\definecolor{originlightgrey}{HTML}{C0C0C0}
\definecolor{origingrey}{HTML}{9F9F9F}
\definecolor{origingreen}{HTML}{10C73E}
\definecolor{corelsteelblue}{HTML}{7E71B1}
\definecolor{originorange}{HTML}{FF7900}

\begin{document}

%\preprint{AIP/123-QED}

\title[Multi-type quantum well semiconductor MECSELs]{Multi-type quantum well semiconductor membrane external-cavity surface-emitting lasers (MECSELs) for widely tunable continuous wave operation}% Force line breaks with \\
%\thanks{Footnote to title of article.}

\author{Patrik~Rajala} \email{patrik.rajala@tuni.fi}
\author{Philipp~Tatar-Mathes}
 \affiliation{Optoelectronics Research Centre (ORC), Physics Unit / Photonics, Faculty of Engineering and Natural Science, Tampere University, Korkeakoulunkatu 3, 33720 Tampere, Finland}

\author{Hoy-My~Phung}
\affiliation{Optoelectronics Research Centre (ORC), Physics Unit / Photonics, Faculty of Engineering and Natural Science, Tampere University, Korkeakoulunkatu 3, 33720 Tampere, Finland}
\affiliation{Robert Bosch GmbH, Robert-Bosch-Campus 1, 71272 \mbox{Renningen, Germany}}

\author{Jesse~Koskinen}
\affiliation{Optoelectronics Research Centre (ORC), Physics Unit / Photonics, Faculty of Engineering and Natural Science, Tampere University, Korkeakoulunkatu 3, 33720 Tampere, Finland}
\author{Sanna~Ranta}
\affiliation{Optoelectronics Research Centre (ORC), Physics Unit / Photonics, Faculty of Engineering and Natural Science, Tampere University, Korkeakoulunkatu 3, 33720 Tampere, Finland}
\author{Mircea~Guina}
\affiliation{Optoelectronics Research Centre (ORC), Physics Unit / Photonics, Faculty of Engineering and Natural Science, Tampere University, Korkeakoulunkatu 3, 33720 Tampere, Finland}

\author{Hermann~Kahle} \email{hermann.kahle@uni-paderborn.de}
\affiliation{Optoelectronics Research Centre (ORC), Physics Unit / Photonics, Faculty of Engineering and Natural Science, Tampere University, Korkeakoulunkatu 3, 33720 Tampere, Finland}
\affiliation{Institute for Photonic Quantum Systems (PhoQS), Center for Optoelectronics and Photonics Paderborn, and Department of Physics, Paderborn University, Warburger Straße 100, 33098 \mbox{Paderborn, Germany}}
\date{\today}% It is always \today, today,
             %  but any date may be explicitly specified

\begin{abstract}
Membrane external-cavity surface-emitting lasers (MECSELs) are at the forefront of pushing the performance limits of vertically emitting semiconductor lasers. Their simple idea of using just a very thin (hundreds of nanometers to few microns) gain membrane opens up new possibilities through uniform double side optical pumping and superior heat extraction from the active area. Moreover, these advantages of MECSELs enable more complex band gap engineering possibilities for the active region by the introduction of multiple types of quantum wells (QWs) to a single laser gain structure. In this paper, we present a new design strategy for laser gain structures with several types of QWs. The aim is to achieve broadband gain with relatively high power operation and potentially a flat spectral tuning range. The emphasis in our design is on ensuring sufficient gain over a wide wavelength range, having uniform pump absorption, and restricted carrier mobility between the different quantum wells during laser operation. A full-width half-maximum tuning range of $>$\,70\,nm ($>$\,21.7\,THz) with more than 125\,mW of power through the entire tuning range at room temperature is demonstrated.
\end{abstract}

%Use showkeys class option if keyword
                              %display desired
\maketitle
\begin{quotation}
During the last decades many application fields ranging from atomic and molecular physics\cite{Wieman.Hollberg_1991} to spectroscopy\cite{Chenais.ea_2002,Bertinetto.ea_1995} and from broadband sensors to optical telecommunication systems\cite{Coldren.Fish.ea_2004,Berger.Anthon_2003} have shown an increasing demand for widely tunable {lasers\cite{Coldren_2000}}. Due to their compactness, low costs, and relatively high efficiencies, semiconductor lasers are in general favoured over other laser systems. When it comes to highly tunable lasers, titanium-sapphire  lasers (Ti:sapph.) are the gold standard. Their ability to produce a flat emission power spectrum on wavelengths ranging roughly from 650\,nm to 1100\,nm is superior to any semiconductor laser \cite{Moulton_1986,Pinto_1994}. However, many application fields mentioned above do not require this whole tuning range or the power level that Ti:sapph.~provides. At the same time, they would benefit strongly from the unique properties of semiconductor lasers, particularly the compactness, ease of manufacturing and lower costs, if these could be achieved while maintaining a relatively flat spectral tuning range. Membrane external-cavity surface-emitting lasers {(MECSELs)\cite{Yang.Albrecht.ea_2015b,Kahle.Mateo.ea_2016a}} are able to answer this need by enabling the use of advanced gain structures with multi-type quantum well (QW) designs, as it will be demonstrated in this paper.
\end{quotation}
\section{\label{sec:introduction}Introduction}

\begin{figure}[htbp]
\centering
\includegraphics[width=0.8\textwidth]{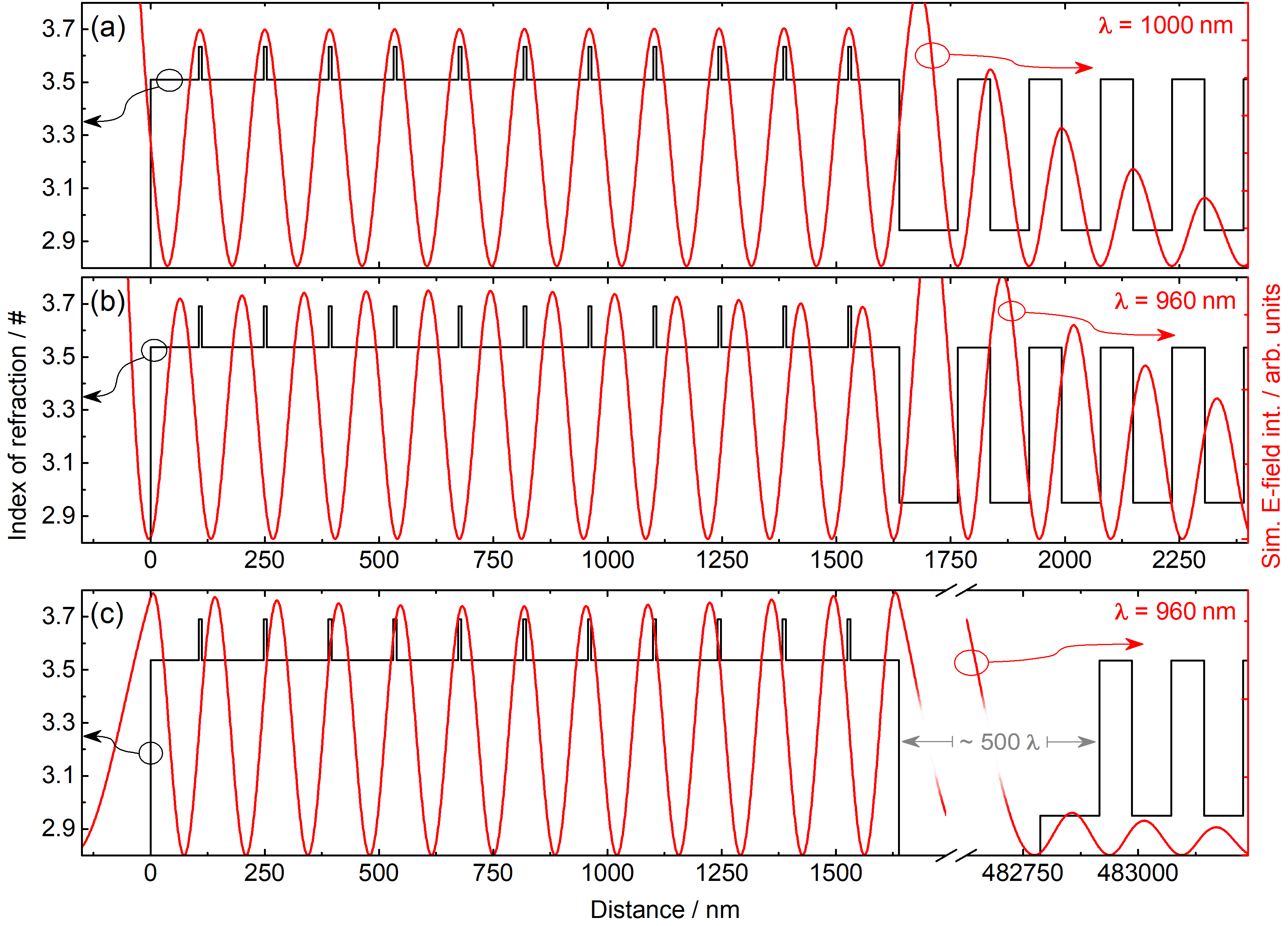}
\caption{\label{fig:phase}(a) A simplified exemplary VECSEL gain chip for operation at 1\,\textmu m wavelength with simulated E-field intensity. The antinodes of the standing wave of the optical field match with the QWs and resonant operation is present. (b) The exact same structure with simulated E-field intensity at 960\,nm. The standing wave of the optical field no longer overlaps with the QWs and the laser emission would stop as no field enhancement could be created. (c) The same active region, but now in a MECSEL configuration with the external cavity mirror about 500\,$\lambda$ away from the active region membrane. The laser ``finds'' the best possible overlap between the standing wave of the optical field and the quantum wells, even though the match is no longer perfect.}
\end{figure}

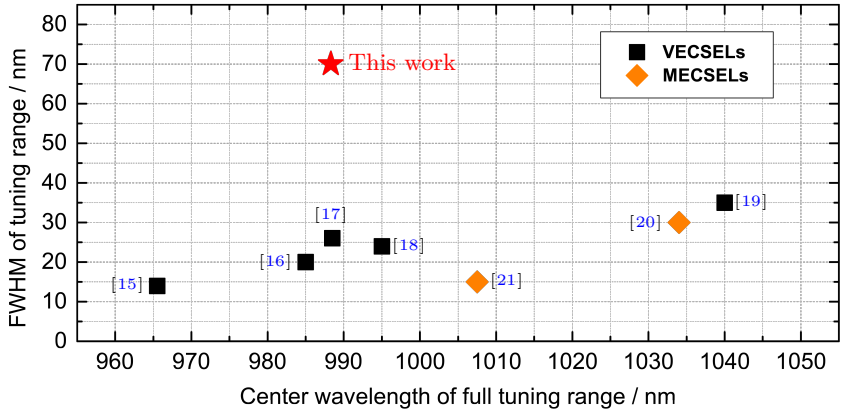
\begin{figure}[ht]
 \centering 
 \includegraphics[width=0.8\textwidth]{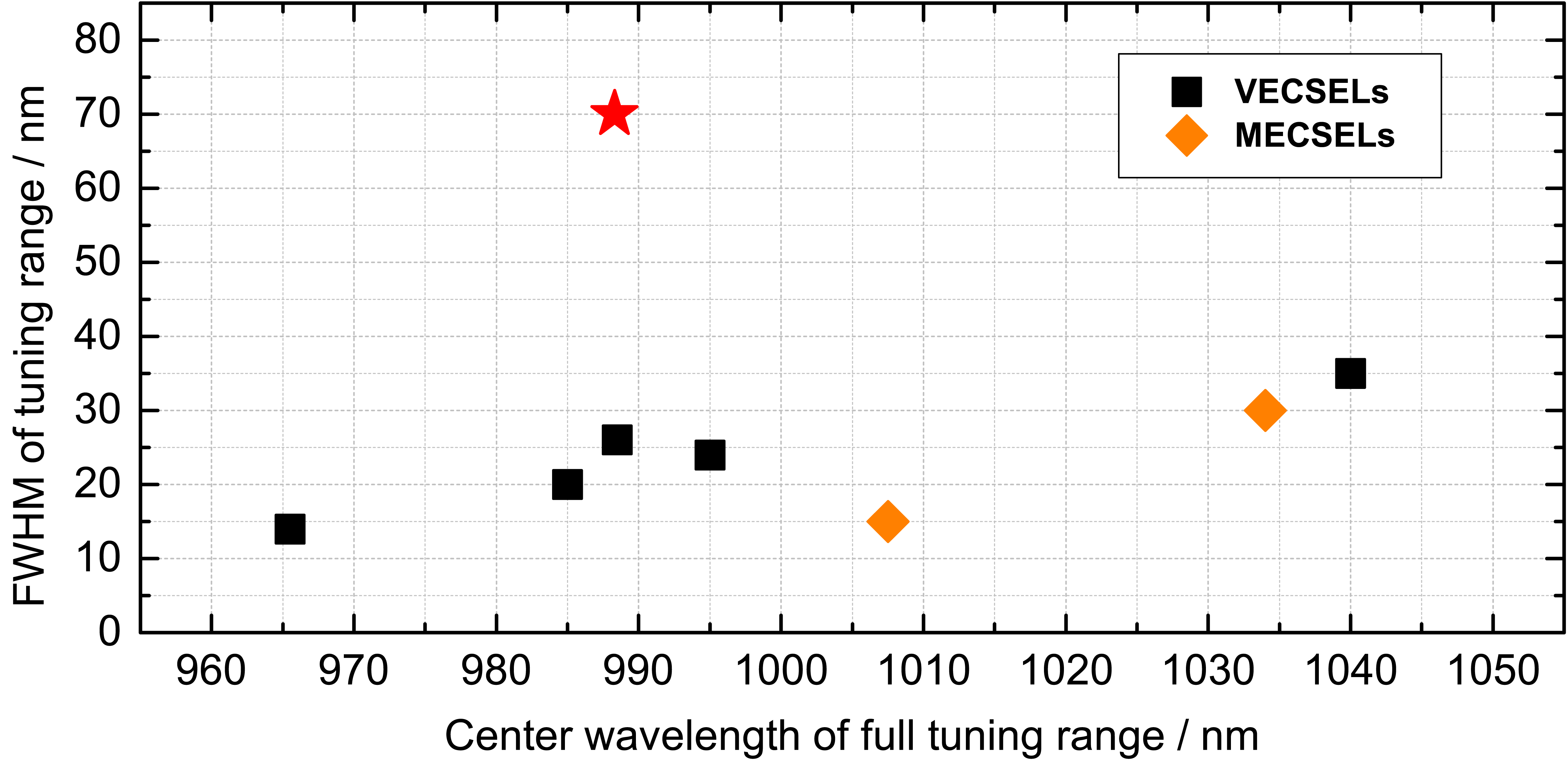}
 %\put(-421,117){\textcolor{black}{$^{[}$\cite{Kahle.Mateo.ea_2016a}}$^{]}$}%661.8 nm, 23.7 nm
%\put(-418,139){\textcolor{black}{$^{[}$\cite{Tatar-Mathes.Kahle.ea_2017}}$^{]}$}%667.5 nm, 25.4 nm
%\put(-380,115){\textcolor{black}{$^{[}$\cite{Schlosser.Hastie.ea_2009}}$^{]}$}%735.0 nm, 26 nm
%\put(-387,146){\textcolor{black}{$^{[}$\cite{Kahle.Nechay.ea_2018}}$^{]}$}%769.0 nm, 40.5 nm
%\put(-359,165){\textcolor{black}{$^{[}$\cite{Kahle.Penttinen.ea_2019}}$^{]}$}%789.25 nm, 44.5 nm
%\put(-342,85){\textcolor{black}{$^{[}$\cite{Phung.Kahle.ea_2020a}}$^{]}$}%825 nm, 22.0 nm
%\put(-347,121){\textcolor{black}{$^{[}$\cite{Hastie.Hopkins.ea_2003}}$^{]}$}%847 nm, 34.0 nm
\put(-276,42){\textcolor{black}{$^{[}$\cite{Mao.Zhang.ea_2021}}$^{]}$}%965.5 nm, 45.0 nm
\put(-220,51){\textcolor{black}{$^{[}$\cite{Broda.WojcikJedlinska.ea_2017}}$^{]}$}%985 nm, 70 nm
\put(-199,68.5){\textcolor{black}{$^{[}$\cite{Borgentun.Bengtsson.ea_2010a}}$^{]}$}%988.5 nm, 43 nm
\put(-169.5,57){\textcolor{black}{$^{[}$\cite{Borgentun.Hessenius.ea_2011}}$^{]}$}%995 nm, 32 nm 
\put(-132.5,43.5){\textcolor{black}{$^{[}$\cite{Mirkhanov.Quarterman.ea_2017}}$^{]}$}%1007.5 nm, 27 nm
\put(-80,65.5){\textcolor{black}{$^{[}$\cite{Yang.Albrecht.ea_2016}}$^{]}$}%1034 nm, 80 nm
\put(-40,73.5){\textcolor{black}{$^{[}$\cite{Butkus.Rautiainen.ea_2011}}$^{]}$}%1040 nm, 60 nm
%\put(-212,142){\textcolor{black}{$^{[}$\cite{Yang.Albrecht.ea_2015b}}$^{]}$}%1152 nm, 78 nm
%\put(-232,123){\textcolor{black}{$^{[}$\cite{Priante.Zhang.ea_2022}}$^{]}$}%1159.5 nm, 71 nm
%\put(-200,117){\textcolor{black}{$^{[}$\cite{Butkus.Rautiainen.ea_2011}}$^{]}$}%1180 nm, 69 nm
%\put(-167,54){\textcolor{black}{$^{[}$\cite{Butkus.Rautiainen.ea_2011}}$^{]}$}%1260 nm, 25 nm
%\put(-95,109){\textcolor{black}{$^{[}$\cite{Saarinen.Lyytikaeinen.ea_2015}}$^{]}$}%1490 nm, 90 nm
%\put(-72,89){\textcolor{black}{$^{[}$\cite{Phung.TatarMathes.ea_2021}}$^{]}$}%1497 nm, 86 nm
%\put(-35,46){\textcolor{black}{$^{[}$\cite{Jezewski.Broda.ea_2020}}$^{]}$}%1640 nm
%\put(-138.0,62){cite{Paajaste.Suomalainen.ea_2009}}
%\put(-137.5,101.5){\cite{Paajaste.Suomalainen.ea_2009}}
%\put(-114.0,63){\cite{Schulz.Hopkins.ea_2008}}
\put(-186,129){\textcolor{red}{This work}}
\caption{Overview over the FWHM of a selection of VECSELs (black squares) and MECSELs (full \textcolor{originorange}{orange} rhombuses) with relatively wide full tuning ranges around 1000\,nm emission wavelength, plotted over their central emission wavelength of their FWHM tuning ranges. Both, quantum well (QW) and quantum dot (QD) based laser active structures are included.}
 \label{fig:overview}
\end{figure}

\textsc{Leinonen}~\textit{et\,al.}\cite{Leinonen.Morozov.ea_2005,Morozov.Leinonen.ea_2006} have demonstrated a vertical external-cavity surface-emitting laser (VECSEL) capable of operating on two different wavelengths at the same time owing to the use of two different kinds of QWs in a single laser gain structure, which proves that the operation of a surface-emitting laser employing a relatively complex QW gain structure is certainly feasible. Wider tunability and operation on two different wavelengths with the aid of utilizing non-linear optical processes, namely second-harmonic generation (SHG) and sum-frequency generation (SFG), has been demonstrated more recently as well \cite{Lukowski_2015}. Other strategies to achieve wide tuning ranges with VECSELs have included a multi chip cavity configuration\cite{Fan.Fallahi.ea_2007}, the use of an inserted blade\cite{Mao.Zhang.ea_2021}, a two-mode resonant microcavity\cite{Broda.WojcikJedlinska.ea_2017}, careful gain element reflectance engineering and optimization of the laser gain element\cite{Borgentun.Bengtsson.ea_2010a,Borgentun.Hessenius.ea_2011} as well as the use of quantum dots (QDs) for quantum confinement\cite{Butkus.Rautiainen.ea_2011}. However, the VECSEL technology used in all of these previous demonstrations imposes certain limits on continuous tuning throughout the wavelength range enabled by the QWs or QDs. The main limitation arises from the monolithically integrated distributed \textsc{Bragg} reflector (DBR), which locks the phase of the standing wave of the optical field in a VECSEL's gain region during laser operation, as can be seen in Fig.\,\ref{fig:phase}.
Figure\,\ref{fig:phase}a demonstrates how the standing wave of the optical field in a VECSEL on the designed operation wavelength, which in our example is 1\,{\textmu}m, overlaps well with the QWs to create field enhancement. Figure\,\ref{fig:phase}b in turn demonstrates how changing the operation wavelength away from the design wavelength (demonstrated in Fig.\,\ref{fig:phase}a shifts the overlap between the standing wave of the optical field and the gain providing QWs completely out of phase, because the DBR locks the phase of the optical field at its interface. In MECSELs, this limitation shown in Fig.\,\ref{fig:phase}b is removed \cite{Yang.Albrecht.ea_2016}, as both mirrors are external and the phase of the optical field on the edges of the laser membrane is not pre-determined, but rather depends on the current state of the external mirrors and the entire cavity. This gives the laser the ability to ``find'' the best possible overlap between the standing wave of the optical field and the gain providing QWs in any situation and provide enough gain for laser operation even on a wavelength that is far from the designed operation wavelength, as is illustrated in Fig.\,\ref{fig:phase}c. The chosen wavelength range in this paper is around 1\,{\textmu}m, because several notable benchmarks for wide tuning range exist in this wavelength range and because the material system utilized is very well developed and thus does not introduce any extra variables. Figure\,\ref{fig:overview} summarizes the development of tuning bandwidths in the 1\,{\textmu}m wavelength range by showing the advances made with VECSELs and MECSELs and how the work presented in this paper compares to them. The comparison here is made by considering the full-width half-maximum (FWHM) of the emission spectrum, as that is what is in the end significant, if we are to eventually have a constant power over the tuning range resulting in a  flat emission spectrum. It is worth noting that the VECSELs seen in Fig.\,\ref{fig:overview} are explicitly designed for wide tuning, while the MECSELs are not (due to their relative novelty in the field), which already shows the potential and inherent advantage when it comes to tuning range, as demonstrated with Fig.\,\ref{fig:phase}. We can also see from Fig.\,\ref{fig:overview} that when it comes to FWHM of the tuning range, our design utilizing multiple types of QWs is a highly effective strategy that does not rely on complicated processing, coatings or pumping setups, but just optimized epitaxial design of the gain structure.\\

\section{\label{sec:designstructuresim}Design and structure simulation}

\begin{figure}[htbp]
\centering
\subfloat[]{\includegraphics[width=0.49\textwidth]{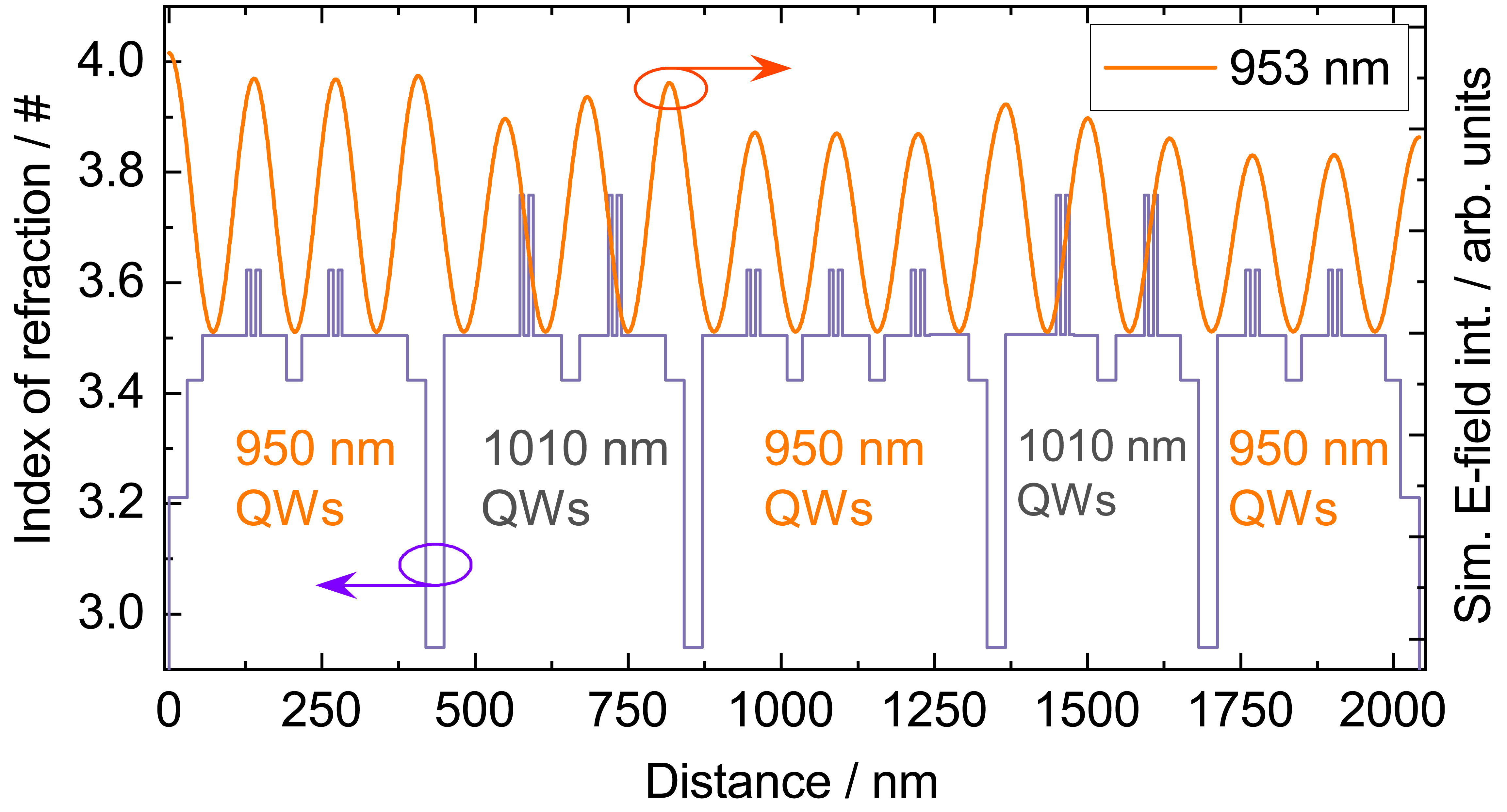}%
\label{fig:953}}
\hfill
\subfloat[]{\includegraphics[width=0.49\textwidth]{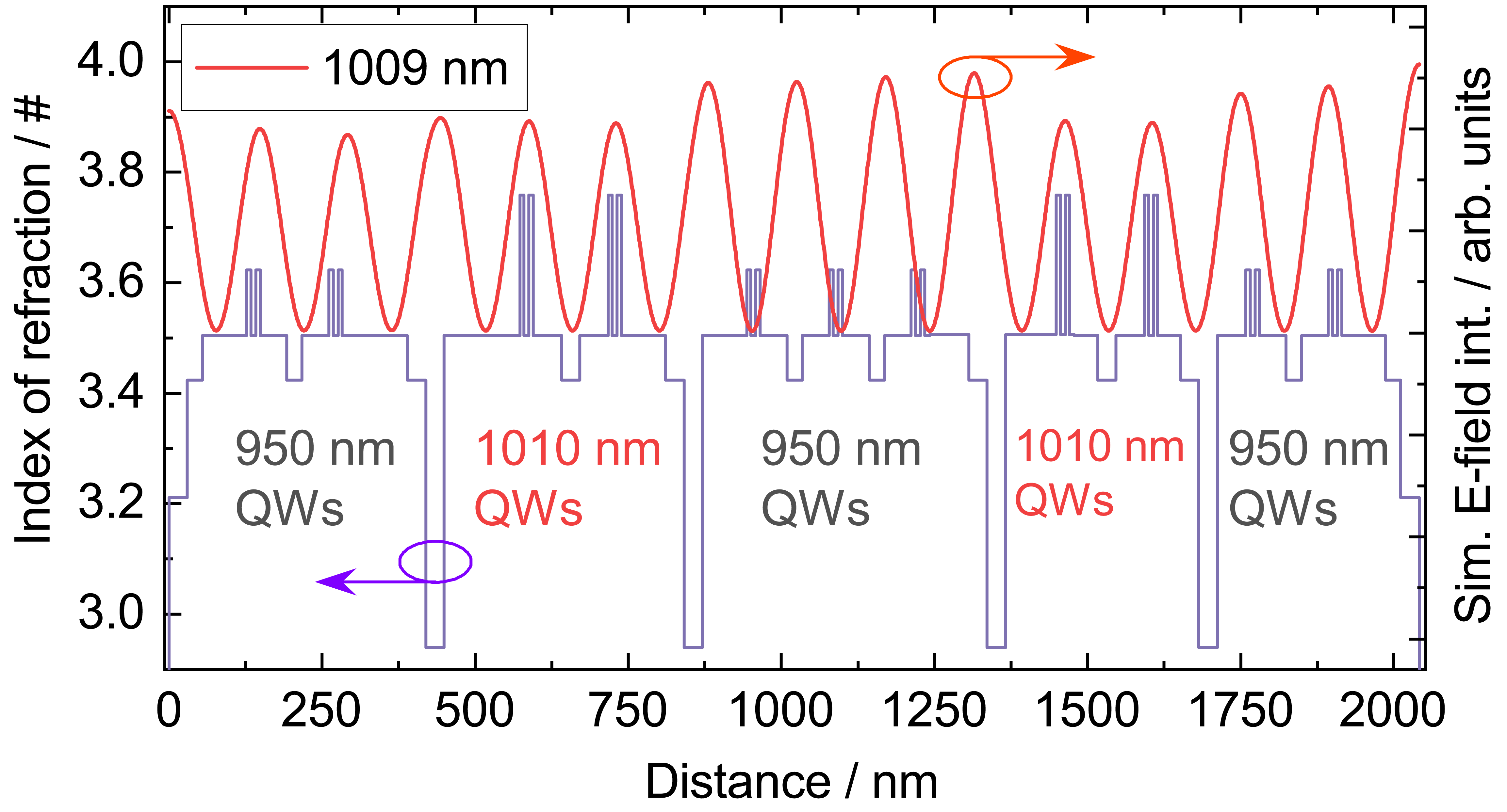}%
\label{fig:1009}}\\
\vspace{-0.2cm}
\caption{\label{fig:structuresim} The gain membrane's architecture is shown. (a) Refractive index (\textcolor{corelsteelblue}{violet} solid line) and the simulated standing wave E-field intensity (solid \textcolor{originorange}{orange} line) of the short wavelength (nominally 953\,nm) resonance are plotted over the thickness of the gain membrane. (b) Refractive index (\textcolor{corelsteelblue}{violet} solid line) and the simulated standing wave E-field intensity (solid \textcolor{originred}{red} line) of the long wavelength (nominally 1009\,nm) resonance are plotted over the thickness of the gain membrane.}
\end{figure}

While the addition of multiple types of QWs to a MECSEL gain structure does not significantly complicate the epitaxial growth or the processing of the MECSEL, there are few important additional steps in the design phase that need to be taken into account. In the first demonstration of a multi-type QW surface-emitting laser by \textsc{Leinonen}~\textit{et\,al.}\cite{Leinonen.Morozov.ea_2005}, there were two main design ideas in this dual-wavelength VECSEL: (1) separating the different kinds of QWs from each other with electron blocking layers (EBLs) to prevent most of the excited charge carriers from diffusing extensively towards the lowest energy states inside the structure, i.e.~the QWs of the longer emission wavelength, and (2) matching the standing wave of the electric field during operation in a way that ensures that when the laser is operated on the shorter wavelength, QWs corresponding to that emission wavelength are in the nodes of the electric field and the other QWs are in the antinodes, and vice versa for the longer wavelength operation. Moving from dual-wavelength operation to continuous broadband tuning operation with a narrow wavelength range at a time, design point (1) is still valid, as the charge carriers will otherwise always diffuse mostly towards the energy states of the longer wavelength QWs. This would significantly reduce the tuning range available, which was effectively demonstrated by the photoluminescence (PL) measurement results by \textsc{Leinonen}~\textit{et\,al.}\cite{Leinonen.Morozov.ea_2005}. Point (2) is not as straightforward for the cases of continuous tuning and using MECSEL technology instead of VECSELs. It is clear that on the shorter wavelength end it is beneficial to have clear matching of antinodes on the shorter emission wavelength QWs and matching of nodes on the longer emission wavelength QWs to minimize the re-absorption by the long wavelength QWs during operation. However, the opposite way this should not play a role, as the QWs of shorter emission wavelength are not capable of absorbing photons emitted by the QWs of longer emission wavelength. The absence of a phase-locking DBR allows the laser to choose the suitable phase on whatever wavelength it is currently operating (see also \textsc{Tatar-Mathes}~\textit{et\,al.}\cite{TatarMathes.Phung.ea_2023}). Thus it is more important to just have a good matching of antinodes on the QWs of the longer emission wavelength, when we are operating the MECSEL on the longer side of the wavelength range.
\\
In addition to the aspects mentioned, the choice of QW materials, their number both per group and in total, and the consideration of pump power distribution inside the structure during operation, are of high importance. For the choice of QW materials, we must consider the tunability of a single type of QW. For VECSELs consisting of a single type of QW, a wide, full tuning range in this wavelength range has been demonstrated for example by \textsc{Borgentun}~\textit{et\,al.}\cite{Borgentun.Hessenius.ea_2011,Borgentun.Bengtsson.ea_2010a} with 32\,nm and 43\,nm, respectively, and by \textsc{Broda}~\textit{et\,al.}\cite{Broda.WojcikJedlinska.ea_2017} with 70\,nm (without temperature tuning). For MECSELs the latest record values were published by \textsc{Yang}~\textit{et\,al.} \cite{Yang.Albrecht.ea_2016, Yang.Albrecht.ea_2015b} and \textsc{Priante}~\textit{et\,al.} \cite{Priante.Zhang.ea_2022} within the 1.0\,\textmu m to 1.2\,\textmu m range. As seen from these results, the tuning range is not evenly distributed around the center wavelength, but follows the typical inverted parabolic behaviour, which reassembles the spectral shape of the gain above laser threshold (if the cavity mirrors possess a flat reflection characteristic). Thus the choice of QW materials is a trade-off between maximum tuning range and constant emission power everywhere within, when using multiple QW types in a single structure. In order to receive a relatively flat emission power distribution throughout the tuning range, based on previous experiments, a separation of 60\,nm in nominal emission wavelengths was chosen. Suitable nominal wavelengths in the chosen 1\,{\textmu}m wavelength range were then 950\,nm and 1010\,nm. These were achieved by using 7\,nm thick InGaAs QWs with an In fraction of 13.0\,\% and 20.5\,\%, respectively. In addition, GaAs barriers, GaAsP strain compensation layers, AlAs EBLs (transparent to both the pump and emission wavelengths), and lattice matched GaInP window layers were used in the structure. The number of QWs per group that can be used, is almost entirely dictated by the amount of local strain that the structure can withstand and thus an amount of two QWs per QW group was chosen in our design. 
%The In fractions of the QWs used introduce enough strain that not more than 2 QWs per QW group could be utilized, as a strain compensation layer is then needed to alleviate and stop the local strain from accumulating over the general estimate of 20 \%nm\cite{Ohtake_2022,Adams_2011}.
\\
The pump power distribution during operation was taken into account when deciding the number and distribution of the chosen QWs inside the structure. QWs of the shorter emission wavelength (950\,nm) have 37\,\% less quantum confinement for electrons and 33\,\% for holes compared to the 1010\,nm QWs. Thus they provide less small-signal gain. In order to have a (more) homogeneous emission characteristic throughout the tuning range, more of the short wavelength QWs must be utilized than of those of longer nominal emission wavelength. Based on the simulations by \textsc{Phung}~\textit{et\,al.}\cite{Phung.Tatar-Mathes.ea_2022}, we were able to derive that $\sim$\,2\,{\textmu}m is a good value of total thickness for the structure, so that we are still able to pump all QWs sufficiently with single-side pumping (SSP) under the given conditions. The structure otherwise was designed to be double-side pumped (DSP) (i.e.~the distribution of QWs), but for comparison we wanted the structure to be able to operate with SSP as well, to be able to compare the changes between the different pumping regimes. Thus we decided to grow the 2\,{\textmu}m thick structure, which allows us to use eleven QW groups of two QWs per group, when we take into account how much spacer, barrier, charge carrier blocking, and strain compensation layers are also needed. In order to match the $\sim$\,35\,\% difference in quantum confinement between the two types of QWs as closely as possible, a distribution of seven short and four long wavelength QW packages was chosen.
\\
When the amounts and specific materials of the layers as well as the total thickness of the structure are decided, the only thing left to do is to place all of them inside the structure accordingly. Achieving a more homogeneous pump power distribution during DSP, the QW distribution was chosen to be symmetric when looking from the middle of the structure. Due to antinode and node matching the structure itself in detail is not completely symmetric as the placements and thicknesses of the spacer layers were adjusted last in the design process. This was done by simultaneously matching antinodes on QWs of the shorter wavelength and nodes on the QWs of the longer wavelength on the shorter target wavelength of 950\,nm. The same was done for the antinodes on the QWs of the longer wavelength on the longer target wavelength of 1010\,nm. Figures\,\ref{fig:953} and \ref{fig:1009} show the simulated standing wave of the electric field intensity inside the structure during operation on wavelengths of 953\,nm and 1009\,nm, respectively. The structure was designed using the SimuLase\textsuperscript{\textregistered} software and the optical field simulations were made using the transfer matrix method in the expected lasing temperature of 350\,K. In addition to the node and antinode matching described above, Figs.\,\ref{fig:953} and \ref{fig:1009} show the ordering of the QW groups in five sections divided by the EBLs mentioned at the beginning of this section.
This kind of order and distribution was chosen as a balance between keeping the structure simple enough to have good antinode and node matching on different wavelengths without thickening the structure needlessly, as well as having a rather balanced amount of pump light reaching each of the QW group sections in both SSP and DSP situations. As a consequence of this trade-off, three antinodes (at distances $\sim$\,425\,nm, 850\,nm, and 1350\,nm, see Figs.\,\ref{fig:953} and \ref{fig:1009}) could not be brought in overlap with any QW pair as the space was needed for the EBLs. Based on the \textsc{Phung}~\textit{et\,al.}\cite{Phung.Tatar-Mathes.ea_2022} simulation, we could estimate that 62\,\% of the pump light is absorbed in the shorter wavelength QW group sections and 33\,\% in the longer wavelength QW group sections in the structure depicted in Figs.\,\ref{fig:953} and \ref{fig:1009}. Considering the seven to four relation in the number of QW groups, this is a very good balance in terms of QWs receiving an even amount of pump light, when finally applying DSP.
\\
The sample was mounted in the lasing setup shown in Fig.\,\ref{fig:setup} in the same orientation as it is depicted in Figs.\,\ref{fig:953} and \ref{fig:1009}. Therefore, when applying SSP, a slight benefit for right side pumping is likely, as the barriers around the QWs are thicker on the left side of the structure. This might compensate the smaller pump power density on the left side of the gain membrane, when pumping from the right. A more homogeneous pumping condition for all QWs would be the consequence, which in turn would lead to better performance. This is further discussed in the characterization results in section "Broadband wavelength tuning".

\subsection{\label{subsec:pl}Photoluminescence characterization}

The structure was grown by molecular beam epitaxy with a VG\,V80 machine. While the growth parameters of all of the semiconductor materials used in this structure are known very well, the usage of more than one kind of QW in a single structure requires some uncommon calibration methods. The method of choice applied here was to first grow PL samples for each kind of QW separately, fine tune their material composition, and then combine them in the final structure. 
\begin{figure}[htbp]
\centering
\includegraphics[width=1.0\textwidth]{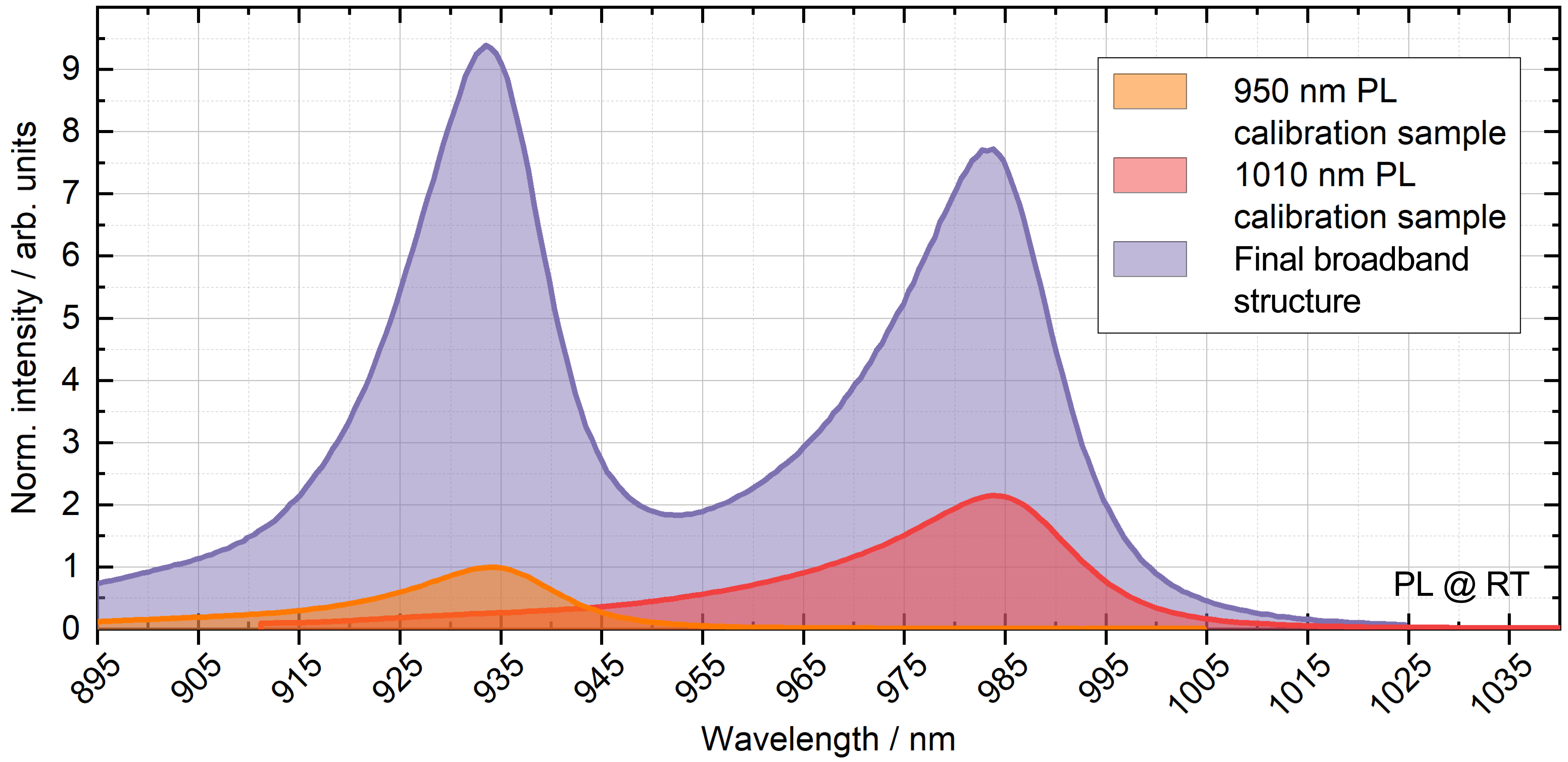}
\caption{\label{fig:pl}The PL signal at room temperature of the final broadband laser structure (\textcolor{corelsteelblue}{violet}) compared to the 950\,nm QW calibration sample's (\textcolor{originorange}{orange}) and 1010\,nm QW calibration sample's (\textcolor{originred}{red}) PL. Measurement was done with an RPM 2000 system using a 785\,nm continuous wave excitation laser. The intensities are normalized to the 950\,nm QW calibration sample's intensity.}
\end{figure}
The PL calibration samples contain only one barrier and one QW group with two QWs of the same kind in it. Figure\,\ref{fig:pl} shows the measured PL curves of both of the PL calibration samples as well as the final structure's PL curve. There we can see that the PL signal of the final structure is very much as expected based on the PL signals from the two separate QW calibration samples. The peak PL emission wavelengths are very close to what was desired, as we expect the gain maximum to be 15 to 25\,nm red-shifted during laser operation. A much higher intensity of the PL signal of the final device was also expected, as there are a total of eleven QW groups compared to the one of the calibration samples. If we look at the calibration samples' PL intensity compared to each other, we can also see the difference in quantum confinement mentioned in the last chapter, as the 1010 nm QWs emit an almost twice as strong PL signal. However, this effect has been quite well balanced with the QW distribution, as in the final device the relative difference between the PL signal from the two peak wavelengths is much smaller and actually the signal from the shorter wavelength QWs is slightly better. This also allows us to assume that the EBLs are working as intended and prevent the longer wavelength QWs from attracting the vast majority of all charge carriers, when comparing this result to the PL results shown by {\textsc{Leinonen}~\textit{et\,al.} \cite{Leinonen.Morozov.ea_2005}}.

\newpage
\section{Laser characterization}
\label{sec:characterization}

In the following, the characterization setup and measurements are described in detail. All presented results in this section have been performed at the same heat sink temperature of 20$^{\circ}$C. Also, the same laser cavity mirrors and the same cavity configuration have been used for all measurements.

\subsection{Experimentation setup}
\label{subsec:setup}
\begin{figure}[htbp]
 \centering 
 \includegraphics[width=0.9\textwidth]{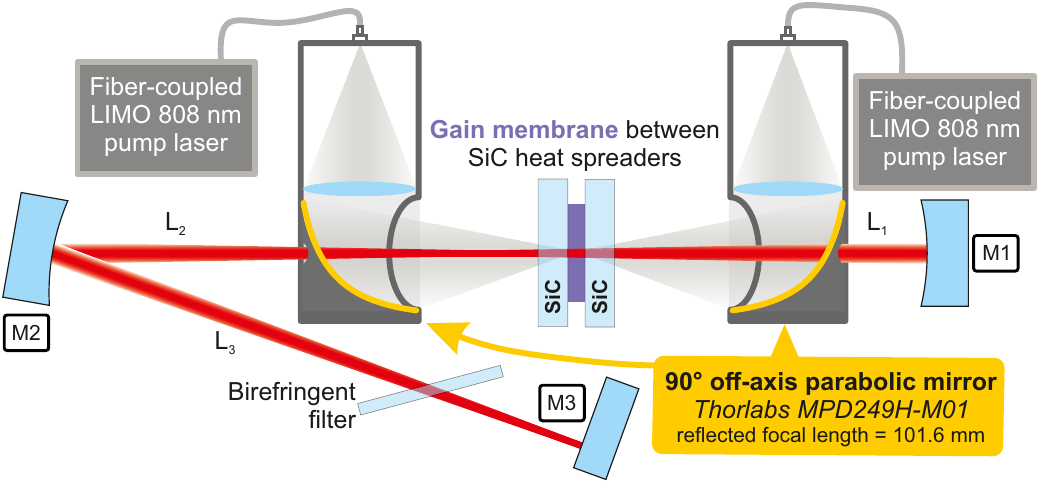}
 \caption{\label{fig:setup}Experimental setup of the MECSEL employing the broadband gain structure and a V-cavity. By using 90$^{\circ}$ off-axis parabolic mirrors with a high-reﬂection protected gold coating, the pump beam is focused down to a nearly circular pump spot onto the laser-active membrane as illustrated.}
\end{figure}

\begin{figure}[htbp]
\centering
\subfloat[]{\includegraphics[width=0.48\textwidth]{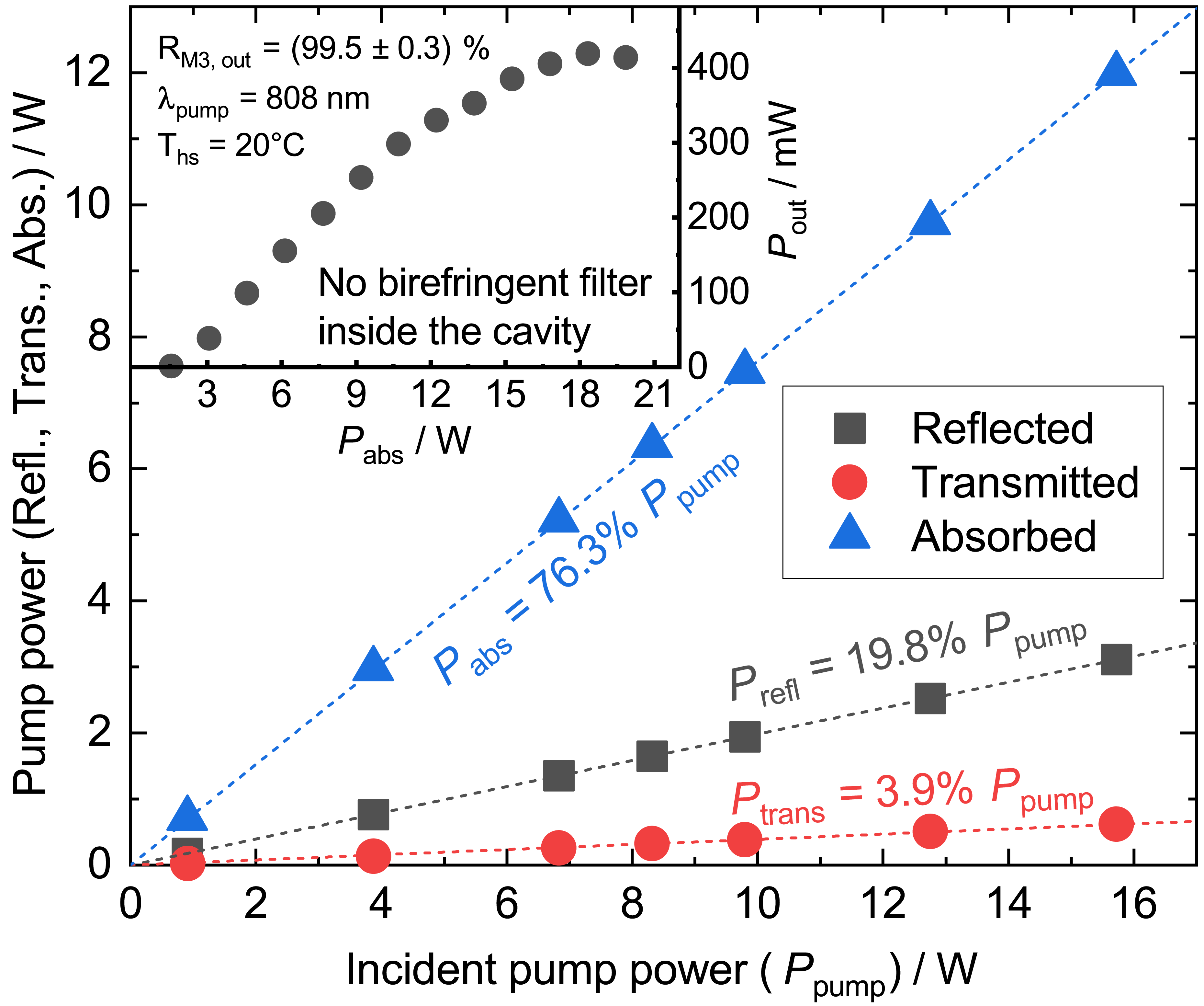}
 %\put(-159.0,34.5){\textcolor{black}{\footnotesize(Fig.\,\ref{fig:spec1})}}%
 %\put(-96.0,110){\textcolor{black}{\footnotesize(Fig.\,\ref{fig:spec2})}}%
 %\put(-28,148){\textcolor{black}{$\uparrow$}}
 %\put(-39.4,139){\textcolor{black}{\footnotesize(Fig.\,\ref{fig:spec3})}}%
\label{fig:powertransfer}}
\hfill
\subfloat[]{\includegraphics[width=0.48\textwidth]{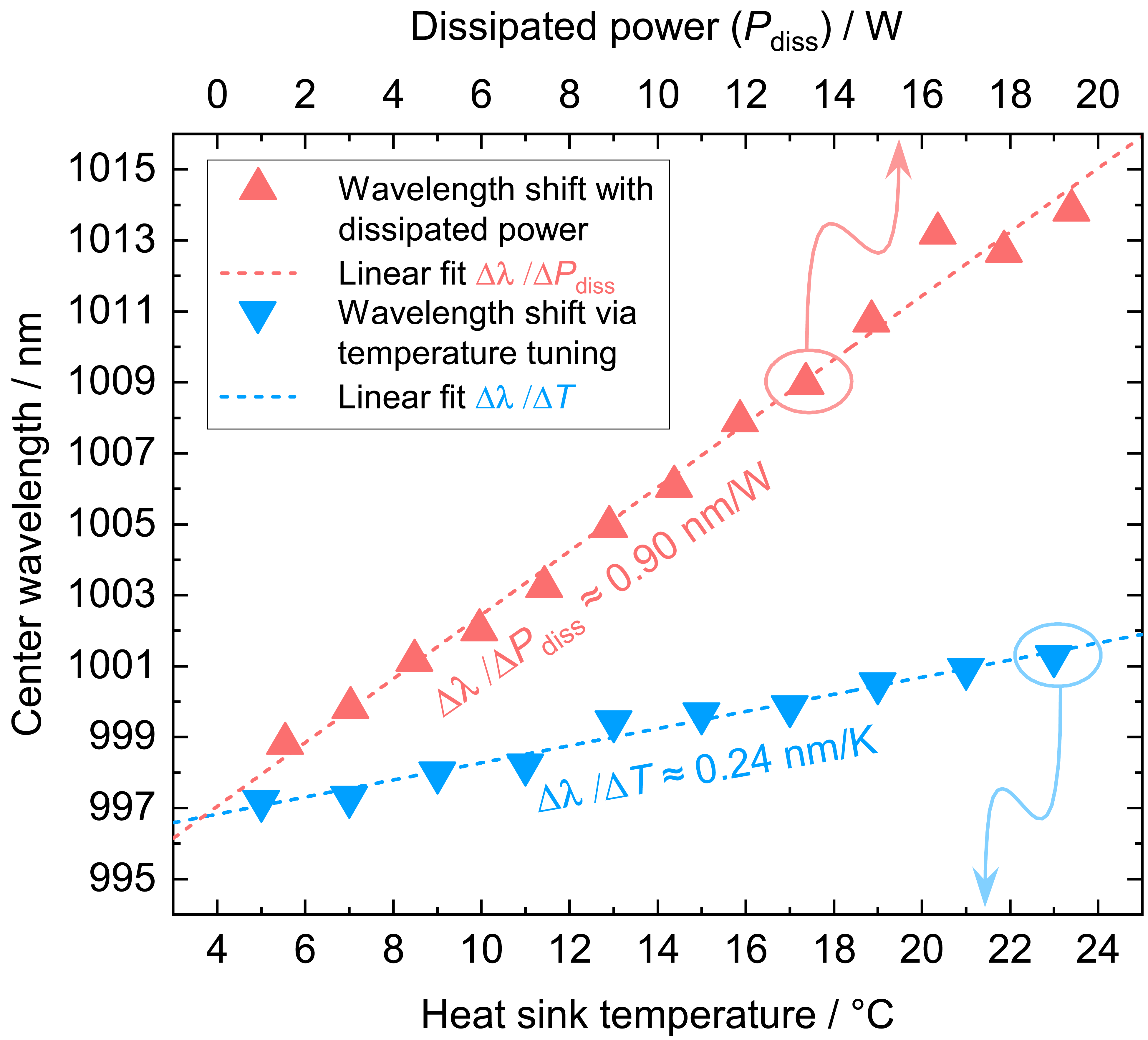}%
\label{fig:dissshift}}\\
\vspace{-0.2cm}
\caption{(a) Distribution of incident pump power $P_{\mathrm{pump}}$ to reflected, transmitted, and absorbed power. (inset) Power transfer measurement under DSP condition to determine the spectral shift $\Delta\lambda$ per absorbed pump power $P_{\mathrm{abs}}$. (b) The power depending spectral shift plotted here was determined during the power transfer measurement plotted in the inset in Fig.\,\ref{fig:powertransfer}.}
\end{figure}

A schematic illustration of the characterization setup is shown in Fig.\,\ref{fig:setup}. The V-cavity consisted of a plane outcoupling mirror M3 with a reflectivity of \mbox{$R_{\text{M3}} = \left(99.5\pm0.3\right)$\,\%}, and two curved high-reflecting mirrors M1 and M2, which both had the same reflectivity of $R_{\text{M1, M2}} > 99.9\,\%$ and the same radius of curvature of \mbox{$r_{\text{M1, M2}} = 300\,\text{mm}$}. The mirror distances of M1 and M2 to the gain membrane sandwich were adjusted to \mbox{$L_1 = 291\,\text{mm}$ and $L_2 = 293\,\text{mm}$}. M3 was positioned under an opening angle of 11$^\circ$ between $L_{2}$ and $L_{3}$. The distance between M2 and M3 was $L_3\,=\,291\,\text{mm}$. For tuning, a 1\,mm thick birefringent filter was used, inserted in the $L_{3}$ cavity arm in an orientation to support parallel polarization. The free spectral range of the birefringent filter at 988\,nm was calculated to be 111.2\,nm. The calculated cavity mode diameter on the gain membrane was about 250\,{\textmu}m using the ray transfer method for a \textsc{Gauss}ian TEM$_{00}$ beam.\\
Two identical 808\,nm LIMO diode lasers (only two construction numbers apart from each other) coupled into identical multi-mode fibers $\left(\mathrm{MHP200L02~from~THORLABS\textsuperscript{\textregistered}}\right)$ with a 200\,{\textmu}m core diameter and a numerical aperture of 0.22 were used as a pump source. 
The fiber output was collimated by a \mbox{$f\,=\,100\,\text{mm}$} anti-reflection coated plano-convex lens, and focused onto a spot size of about 330\,{\textmu}m in diameter by a 90$^\circ$ off-axis parabolic mirror $\left(\mathrm{MPD249H-M01~from~THORLABS\textsuperscript{\textregistered}}\right)$ with a protected gold reflection coating ($R_{808\,\text{nm}}>96\,\%$) and a reflected focal length of 101.6\,mm. Thus, the ratio between the cavity mode and pump spot diameter was about 0.76, which is within the suggested optimum from 0.65 to 0.82 derived by \textsc{Laurain}~\textit{et\,al.}\cite{Laurain.Hader.ea_2019}~via simulations. 
%%%%%%%%%%%%%%%%%%%%%%%
%checked until here, Hermann, 11.01.2023
The parabolic mirror used in these experiments has a diameter of 50.8\,mm as well as a hole with a diameter of 3\,mm that does not cut the intra cavity laser mode. The collimated pump beam irradiated on the parabolic mirror covered almost the whole area of this mirror. For the pump beam, the losses caused by this 3\,mm hole were investigated and were below 1\,\%, and therefore negligible. Nevertheless, the incident pump power $P_{\mathrm{pump}}$ was determined via measuring the power reflected by the parabolic mirror. Therefore, all reflection losses as well as the losses introduced by the centered 3-mm-through-hole and the fiber coupling losses were taken into account. With an angle of incidence of the pump laser, that ranges from almost 0$^{\circ}$ to less than 15$^{\circ}$, this pump approach enables a nearly circular pump spot with a sagittal to tangential diameter relation of $D_{\text{p, sag}}$/$D_{\text{p, tan}}\,>\,0.96$ in the focus.
The MECSEL gain membrane sandwiched between transparent SiC heat spreaders, in the following called MECSEL gain element, was mounted between water/glycol cooled copper heat sinks (not drawn in Fig.\,\ref{fig:setup}) with an aperture diameter of 1.5\,mm. The opening angle of the aperture is 60$^{\circ}$ to provide enough space for the incident pump laser (details can be found elsewhere \cite{Kahle.Penttinen.ea_2019}).\\
The thermal resistance $R_{\mathrm{th}}$ represents one of the most important parameters of MECSELs, it allows to compare and to classify results.
In order to calculate $R_{\mathrm{th}}$, one needs to determine reflected pump power $P_{\mathrm{refl}}$, transmitted pump power $P_{\mathrm{trans}}$, and the absorbed pump power $P_{\mathrm{abs}}$. A power transfer measurement has to be performed to receive the wavelength shift $\Delta\lambda$/$\Delta P_{\mathrm{abs}}$ and the corresponding output power $P_{\mathrm{out}}$. Also, a temperature tuning measurement is essential to determine the thermal shift $\Delta\lambda$/$\Delta T_{\mathrm{hs}}$.
In Fig.\,\ref{fig:powertransfer} it is plotted how the fractions of $P_{\mathrm{pump}}$ in percent are distributed when interacting with the gain sandwich. $P_{\mathrm{refl}}$ was calculated. For 808\,nm the refractive index of SiC is $n_{\mathrm{SiC}}$\,=\,2.60 \cite{Singh.Potopowicz.ea_1971}. For angles of incident between 0$^{\circ}$ and 15$^{\circ}$ the reflectivity of the SiC-air interface changes only in the range of 10$^{\mathrm{-5}}$, therefore a constant reflection of 19.2\,\% can be assumed for unpolarized light. The linear fit to the transmitted power values revealed a transmission of $P_{\mathrm{trans}}$\,=\,3.9\,\%. Now, $P_{\mathrm{abs}}$ can be calculated: $P_{\mathrm{abs}}$\,=\,$P_{\mathrm{pump}}$\,-\,$P_{\mathrm{refl}}$\,-\,$P_{\mathrm{trans}}$\,=\,76.3\,\%. The wavelength shift \mbox{$\Delta\lambda$/$P_{\mathrm{diss}}=\left(0.90\pm0.03\right)$\,nm/W} was extracted from spectra taken during the power transfer measurement plotted in the inset in Fig.\,\ref{fig:powertransfer}. Then it was plotted over the dissipated power $P_{\mathrm{diss}}$ in Fig.\,\ref{fig:dissshift}. $P_{\mathrm{diss}}$ was calculated as follows: $P_{\mathrm{diss}}$\,=\,$P_{\mathrm{abs}}$\,-\,$P_{\mathrm{out}}$, where $P_{\mathrm{out}}$ is the output power measured. Due to the relatively high reflectivity of the used outcoupler of \mbox{$R_{\mathrm{M3,\,out}}=\left(99.5\pm0.3\right)\%$} high power emission was not expected. We see a linear increase in output power starting at the threshold of $P_{\mathrm{th}}=1.63$\,W.
A little kink in the linear behaviour can be seen at about eleven watts of absorbed pump power $P_{\mathrm{abs}}$. This might origin from a polarization jump, as no polarization maintaining element like a birefringent filter is implemented - the laser is operating freely.
Besides the change in the slope efficiency, no change or deviation from the typical linear behaviour in the spectral shift (see Fig.\,\ref{fig:dissshift}) is visible. The setup configuration for this measurement was the same as depicted in Fig.\,\ref{fig:setup}. No birefringent filter or any other optical intra cavity element was used for the power transfer measurement. The pump power was increased symmetrically in the double-side configuration.
Finally, the thermal shift of the laser emission $\Delta\lambda$/$\Delta T_{\mathrm{hs}}$ was measured under constant incident pumping of $P_{\mathrm{pump}}$\,=\,6\,W (3\,W from each side). For this measurement the heat sink temperature $T_{\mathrm{hs}}$ was changed gradually from 5$^{\circ}$C to 23$^{\circ}$C by changing the water/glycol cooling temperature. The temperature difference between chiller and heat sink was determined to be in the range of 0.1$^{\circ}$C and is therefore negligible. The thermal wavelength shift of \mbox{$\Delta\lambda$/$\Delta T_{\mathrm{hs}} = \left(0.24\pm0.01\right)$\,nm/K} was determined via a linear fit to the measured data which are plotted in Fig.\,\ref{fig:dissshift}. This allows to calculate the thermal resistance $R_{\mathrm{th}}$ of the MECSEL gain element including error propagation after \cite{Heinen.Zhang.ea_2012} by dividing the spectral shift per $P_{\mathrm{diss}}$ by the spectral shift per change of $T_{\mathrm{hs}}$ to \mbox{$R_{\mathrm{th}}=\left(3.75\pm0.28\right)$\,K/W}. Considering the use of SiC heat spreaders with a thickness of $\sim$\,350\,\textmu m each, this result meets the expected range \cite{Phung.Tatar-Mathes.ea_2022} and is similar to a previous result in \textsc{Phung}\,\textit{et\,al.}\cite{Phung.Kahle.ea_2020a}.

\subsection{Broadband wavelength tuning}
\label{subsec:tuning}
\begin{figure}[ht]
\centering
\subfloat[]{\includegraphics[width=0.49\textwidth]{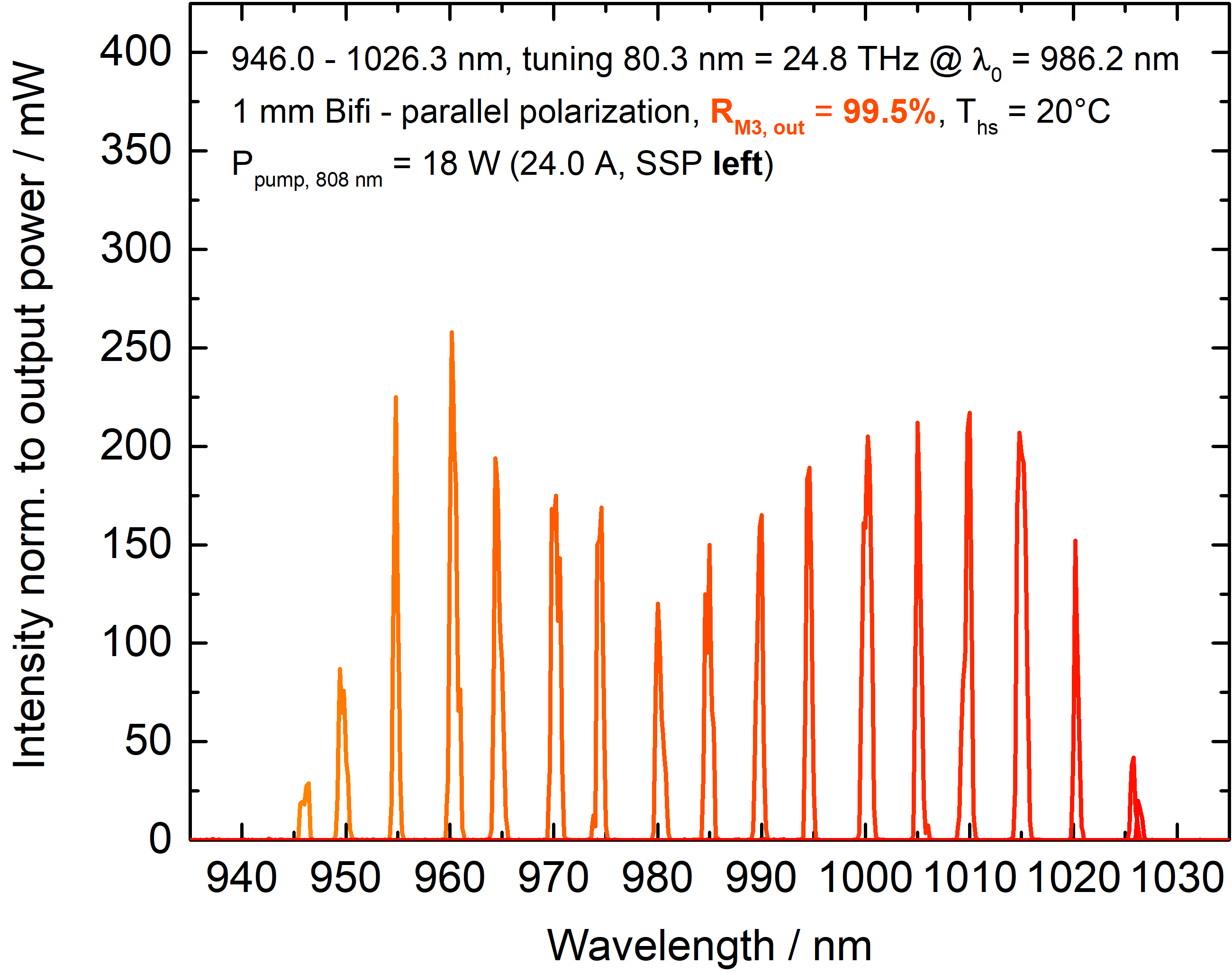}%
\label{fig:SSP-left}}
\hfil
\subfloat[]{\includegraphics[width=0.49\textwidth]{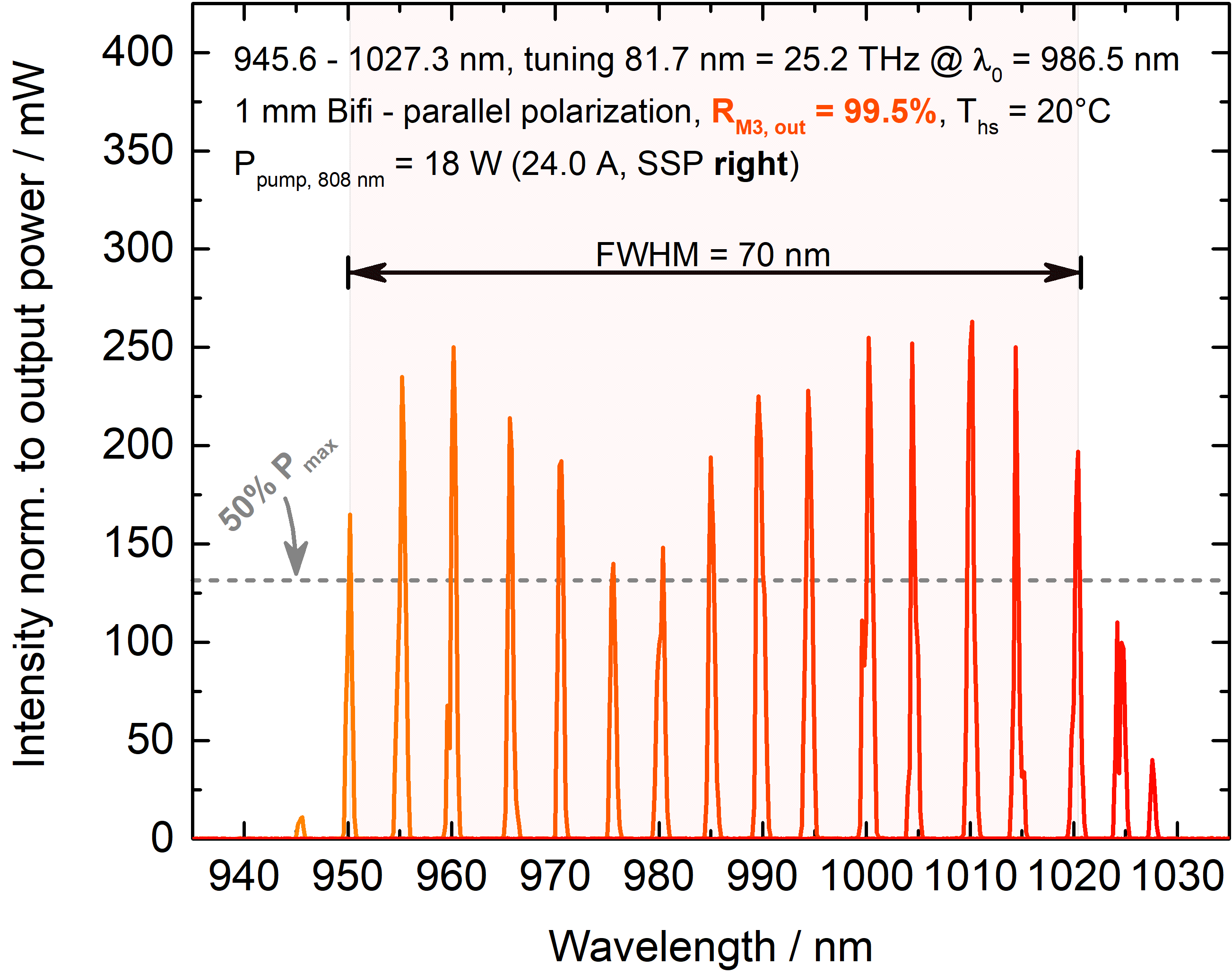}%
\label{fig:SSP-right}}
\vspace{-0.2cm}
\caption{\label{fig:tuningspectra} A set of exemplary spectra (\textcolor{originorange}{orange} $\rightarrow$ \textcolor{red}{red} curves) to depict wavelength tuning are shown in these two plots. The intensity of the spectra was normalized to the correspondingly measured output power, which was plotted over wavelength. 18\,W of pump power were irradiated (a) only from the left, and (b) only from the right side.}
\end{figure}

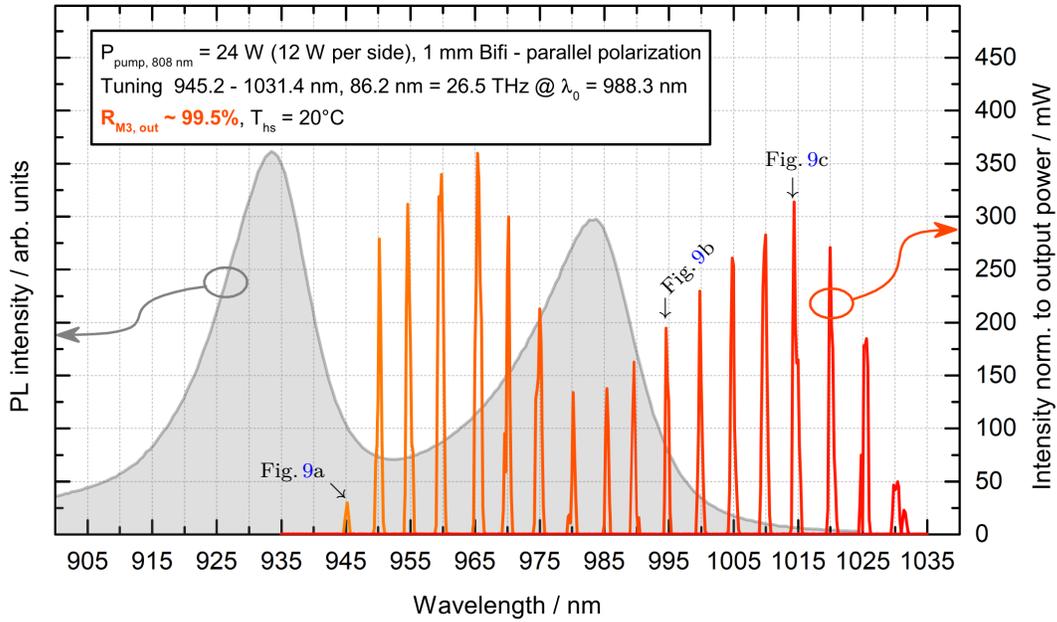
\begin{figure}[ht]
\centering
\includegraphics[width=1.0\textwidth]{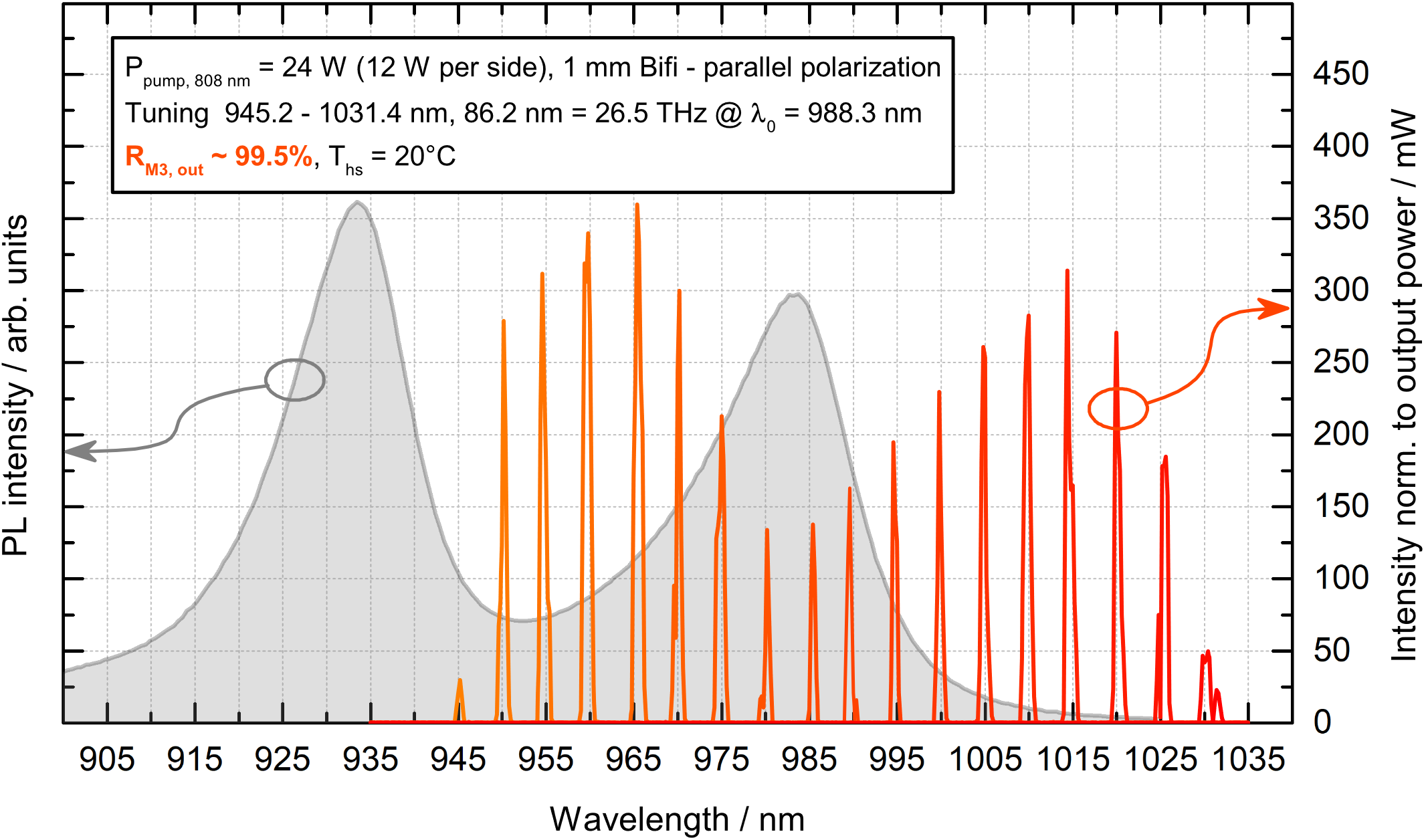}
 \put(-110,172){\textcolor{black}{\footnotesize Fig.\,\ref{fig:beamprofiles}c}}%
 \put(-101.8,161.5){\textcolor{black}{$\downarrow$}}
 \put(-150,123){\textcolor{black}{\rotatebox{45}{\footnotesize Fig.\,\ref{fig:beamprofiles}b}}}%
 \put(-150,116){\textcolor{black}{$\downarrow$}}
 \put(-301,54){\textcolor{black}{\footnotesize Fig.\,\ref{fig:beamprofiles}a}}%
 \put(-276,46){\textcolor{black}{\rotatebox{45}{$\downarrow$}}}
\caption{\label{fig:DSP} A set of exemplary spectra (\textcolor{originorange}{orange} $\rightarrow$ \textcolor{red}{red} curves) to depict wavelength tuning are shown in this plot. The intensity of the spectra was normalized to the correspondingly measured output power, which was plotted over wavelength. Pump power was applied under symmetric DSP condition. The PL (\textcolor{origingrey}{grey} curve with filled area underneath) of the corresponding broadband gain structure from Fig.\,\ref{fig:pl} was plotted for comparison with the tuning behaviour of the MECSEL.}
\end{figure}

For tuning measurements an incident pump power of $P_{\mathrm{pump}}$\,=\,18.0\,W was chosen, because our previous experiments revealed the largest tuning range under SSP at this pump power. Higher pump powers lead to a smaller tuning range and worse absolute output power values due to occurring thermal rollover under this SSP condition. The outer heat spreader facets of the gain element were anti-reflection coated for the wavelengths of the laser as well as the pump to reduce cavity losses. The spectra were recorded with an ANDO\,AQ-6315A{\texttrademark} optical spectrum analyzer (OSA) with 0.05\,nm resolution. The results of SSP measurements are shown in Figs.\,\ref{fig:SSP-left} and \ref{fig:SSP-right}. As described in section ``Design and structure simulation'', we can see a slightly better performance in both output power and tuning range for the SSP case from the right side (Fig.\,\ref{fig:SSP-right}) due to a beneficial barrier width distribution (similar to the structures described for example by \textsc{Phung}~\textit{et\,al.} \cite{Phung.TatarMathes.ea_2021} and \textsc{Baumg\"{a}rtner}~\textit{et\,al.}\cite{Baumgaertner.Kahle.ea_2015}). Additionally, the output power did not drop below 50\,\% of $P_{\mathrm{max}}$ between the performance peaks of the two different QW kinds and a FWHM of 70\,nm was reached maintaining an output power of $>$\,125\,mW.
In the long wavelength range around 1005\,nm the output power is similar to the short wavelength around 960\,nm. For the left side pumping (Fig.\,\ref{fig:SSP-left}) the long wavelength area is about 25\,\% inferior in terms of output power. Although this comparison was not specifically intended, it clearly shows that the QW and barrier width distribution should be matched with the absorption behaviour of the applied pump light in future structure designs. 
In order to obtain a maximum full tuning range, the total incident pump power $P_{\mathrm{pump}}$ was increased to 24\,W and divided equally to 12\,W per side for the case of DSP.
Beforehand experiments (see Fig.\,\ref{fig:powertransfer}) revealed best output power performance and the largest tuning range under DSP condition at this pump power configuration.
As can be seen in Fig.\,\ref{fig:DSP} a wider tuning range of 86.2\,nm or 26.5\,THz was achieved compared to the single-side pumping measurements, where $\sim$\,80\,nm were reached. However, since the peak output powers received at the 965 nm and 1015 nm rose significantly when applying DSP, the FWHM tuning range is no longer as wide, as the output power in the middle of the spectrum is just under half of the peak power. If, however, we look at the tuning range at which 125 mW of output power (which was roughly the FWHM output power level in the SSP case) is maintained, we have an outstanding tuning range of 80\,nm or 25\,THz reaching from 940\,nm to 1020\,nm. 

The PL curve of the corresponding broadband gain structure from Fig.\,\ref{fig:pl} was plotted again in Fig.\,\ref{fig:DSP} for comparison with the laser tuning.
%A further detail which can be seen in Fig.\,\ref{fig:DSP} is that laser emission above 125\,mW (indicated by the \colorbox{originlightred}{light red} background in Fig.\,\ref{fig:DSP}) was achieved in the range of from 950.0 to 1025.0\,nm (75\,nm).
It is further noteworthy how accurately the MECSEL tuning behaviour follows the PL characteristic in a relative intensity comparison taking the spectral red shift and typical inverted parabolic shape of the corresponding QWs' gain into account.

\subsection{\label{subsec:beamprofiles}Beam profiles}

It is a built-in feature of MECSELs to possess the same excellent beam quality properties (M$^{2}$\,<\,1.1 and a TEM$_{00}$ \textsc{Gauss}ian transverse mode profile \cite{Kahle.Mateo.ea_2016a,Phung.Kahle.ea_2020a}) as {VECSELs\cite{Jetter.Michler_2021,Guina.Rantamaeki.ea_2017}}. This is enabled by satisfying the geometrical condition of having the squared gain region thickness $L^2$ much smaller than the laser mode area $s$ ($L^2$\,$\ll$\,$s$) inside the laser cavity\cite{Basov.Bogdankevich.ea_1966}. In this work the relation is 1\,:\,12\,272 with $L^2$\,=\,4\,\textmu m$^2$ and $s$\,=\,49\,087\,\textmu m$^{2}$. The external mirrors provide full mode control as the gain region's impact on beam distortion is present\cite{Phung.Kahle.ea_2020a} but minimal, compared to classical solid-state lasers. Therefore, the behaviour of the transverse mode shape during the tuning measurements is an important characteristic to prove undisturbed operation of the laser. Beam profiles were recorded every second time a spectrum was taken.
\begin{figure}[htbp]
\centering
\includegraphics[width=1.0\textwidth]{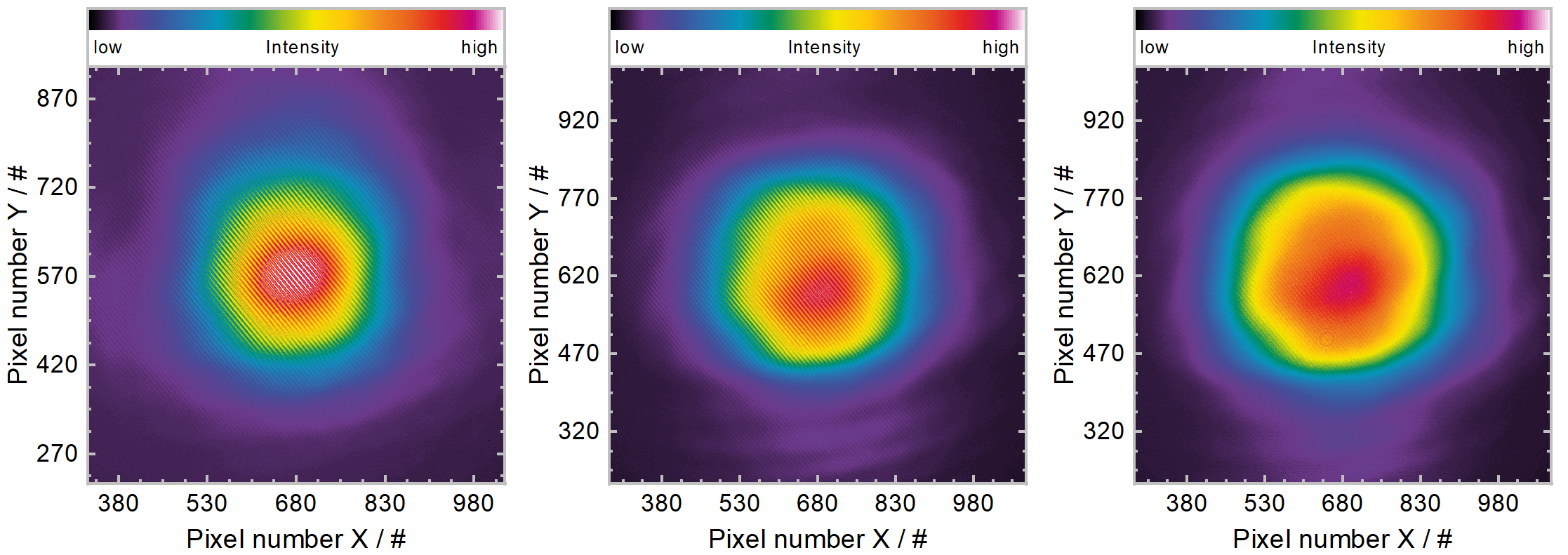}
%\put(-227,24){\textcolor{white}{(a)}}
\put(-368,110){\textcolor{white}{(a)}}
\put(-235,110){\textcolor{white}{(b)}}
\put(-103,110){\textcolor{white}{(c)}}
\caption{\label{fig:beamprofiles}
Beam profiles taken during the tuning measurement plotted in Fig.\,\ref{fig:DSP}. Three examples have been chosen at (a) low ($\sim$\,945\,nm), (b) medium ($\sim$\,995\,nm), and (c) high ($\sim$\,1015\,nm) output power.}
\end{figure}
In Fig.\,\ref{fig:beamprofiles} three exemplary beam profiles for low, medium, and high output power were plotted. In the plot of Fig.\,\ref{fig:DSP} it is indicated to which spectrum the plotted beam profiles of Fig.\,\ref{fig:beamprofiles} corresponds. Small deformation of the beam profile can be seen in Figs.\,\ref{fig:beamprofiles}b and \ref{fig:beamprofiles}c, but the fundamental \textsc{Gauss}ian TEM$_{00}$ character remains intact. During beam profile recording and taking spectra with the OSA, the laser cavity was not realigned and all parameters were kept constant, solely the birefringent filter was rotated. This shows that the excellent beam properties typical to MECSELs were maintained with the novel multi-type QW design.

\section{\label{sec:conclusion}Conclusions}

The widest tuning of a vertically emitting semiconductor laser, 26.5\,THz (86.2\,nm), emitting around 1\,{\textmu}m, was shown. Of greater importance is the FWHM tuning of 70\,nm (and 80\,nm tuning range with steady 125\,mW of output power), which is effectively double to previously reported results in the 1\,{\textmu}m wavelength range. This result shows the tremendous capability of MECSELs utilizing  multiple types of QWs for significantly increasing the tuning range of vertically emitting semiconductor lasers, while maintaining the typical excellent beam quality (M$^{2}$\,<\,1.1\cite{Kahle.Mateo.ea_2016a}) and high power (>\,125\,mW) throughout the whole tuning range. In our design process four relevant strategy points were recognized: 1) separation of different types of QW areas inside the active region from each with EBLs to maintain sufficient carrier concentration in all QWs, 2) ensuring maximum overlap with the standing wave of the optical field on each of the QW types on their designed emission wavelengths, while minimizing the overlap with other QW types, if they are capable of absorbing the photons emitted, 3) choosing correct QW materials and particularly the nominal emission wavelength between them, 4) designing the structure as a whole so that each of the QW types are pumped equally.
\\
\\
This paper has only scratched the surface and shown the applicability when it comes to the potential of multi-type QW MECSELs. The advances that can be made from this point onwards, are two-fold. First of all, more QW types could be utilized to either increase the emission power in the middle of the tuning range if more power throughout the tuning range is needed, or to widen the tuning range by choosing three or more QW types accordingly. Secondly, and more importantly, the total thickness of the MECSEL could be at least doubled to truly take advantage of the MECSELs' double-side pumping capabilities, when we keep in mind that the structure shown in this paper could still be properly pumped from one side only. The doubling of the total thickness allows for much more QW groups to be utilized, which in turn can then be turned into either higher emission power or wider tuning range. In the DSP pumping regime, two pump lasers operating on different wavelengths could also be used to cover a wider pumping wavelength range and avoid absorption in unwanted regions of the active area. Finally, stacking of more than one MECSEL with additional heat spreaders in-between instead of complete monolithic structures is a great option for future widely tunable multi-type QW MECSELs.

\begin{acknowledgments}
The authors would like to thank Antti H\"{a}rk\"{o}nen for the initial, passionate scientific discussions on this project, and gratefully acknowledge the Academy of Finland (No.\,326455), the Academy of Finland \href{https://prein.fi/home}{PREIN} Flagship Programme (No.\,320165), the Horizon 2020 Marie Sk{\l}odowska-Curie Actions - \href{https://netlas.aogkent.uk/}{\mbox{NetLaS}} (No.\,860807), the Magnus Ehrnrooth Foundation, and the Finnish Foundation for Technology Promotion for the financial support.
\end{acknowledgments}

\bibliography{aipsamp}
\end{document}